\documentclass[
reprint,
nofootinbib,
aps, pra
]{revtex4-1}
\usepackage{custom}
\usepackage{comment}
\usepackage{float}
\usepackage{graphicx} 
\usepackage[caption=false]{subfig}
\begin{document}

%========================================
\title{Unruh-DeWitt detector's response to a non-relativistic particle}

\author{Qidong Xu}
\email[]{qidong.xu.gr@dartmouth.edu}
\affiliation{Department of Physics and Astronomy, Dartmouth College, Hanover, New Hampshire 03755, USA}

\date{\today}

\begin{abstract}
Unruh-DeWitt (UDW) detector that couples locally to a quantum field is an important tool for operationally studying field properties. Here, we study the response of an inertial UDW detector to a non-relativistic particle state of a massive scalar field and we find that the transition probability of the detector splits into the vacuum contribution and the matter contribution. We show that the matter part oscillates with the interaction time duration and the oscillation period depends on the difference between the mass of the particle and the energy gap of the detector; a strong resonance pattern for the transition probability is found when such difference equals to zero. We compare the matter part contribution with the non-relativistic probability density of the particle and find that they provide qualitatively similar description of the particle.        
\end{abstract}

\maketitle
%========================================

%========================================
%========================================
\section{Introduction}
A conceptual idealized particle detector model in the context of quantum field theory was initially proposed by Unruh \cite{unruh1976notes} to resolve the ambiguity of defining a physical particle state in a general spacetime background. Later, DeWitt simplified this model by introducing a local two-level system moving along a classical trajectory to replace the field description of the detector \cite{hawking2010general}, which is now known as Unruh-DeWitt (UDW) detector. The UDW detector has a simple interpretation of particles; the transition from the ground state to the excited state of the two-level system is regarded as an absorption of the field quanta, and therefore the detection of a particle of the field.

One of the most well-known example of the UDW detector's application is the proof of the Unruh effect \cite{unruh1976notes}, which states that from the perspective of an uniformly accelerated observer, the Minkowski spacetime vacuum state is a thermal state. As a simple and useful tool, the UDW detector has also received considerable attention in many other areas, including the study of black hole thermodynamics \cite{candelas1977irreversible, davies1978thermodynamics},  Lorentz-violating dispersion relations \cite{gutti2011modified,husain2016low,alkofer2016quantum}, finite spacial extensions of the detector and the corresponding regularization schemes \cite{schlicht2004considerations, langlois2006causal,louko2006often,satz2007then,martin2013wavepacket}, and the coupling to a fermionic field \cite{iyer1980detection,takagi1986vacuum,hummer2016renormalized,louko2016unruh} (for more examples, see recent reviews \cite{crispino2008unruh,hu2012relativistic,martin2014entanglement} and references therein). More recently, UDW detectors have been used extensively in the so-called entanglement harvesting protocol \cite{salton2015acceleration}, where a pair of UDW detectors coupled to a quantum field can be used to extract the vacuum entanglement of the field, and therefore to probe the nontrivial field properties in a wide range of scenarios \cite{reznik2003entanglement,martin2013sustainable, pozas2015harvesting,martin2016spacetime,sachs2017entanglement,ardenghi2018entanglement,trevison2018pure, henderson2020quantum,faure2020particle,ver2009entangling,huang2017dynamics,smithDetectorsReferenceFrames2019,xu2020gravitational,liu2021harvesting}. 

However, despite many successful applications of the UDW detectors, most works primarily focus on the vacuum state of a massless quantum field, with a few exceptions for the massive field and single excitation state \cite{toussaint2021detecting,lee2014spatially}. As a type of particle detector, it is of natural interest to ask how does the UDW detector respond to the field state that represents the matter/particle distribution, and what are the properties of such field state that can be operationally accessed by coupling the field to the UWD detector. Despite the fact that excitation state exhibits quite different theoretical properties including the entanglement entropy \cite{das2006robust} and phase transition \cite{caprio2008excited}, these questions are also directly related to the problem of measuring the quantum field as it is known that the projective measurement does not directly generalize to the framework of quantum field theory \cite{sorkin1993impossible,lin2014notes,borsten2021impossible,bostelmann2021impossible}, while the particle detector based model can be promising to formulate the measurement process \cite{pologomez2021detectorbased}. It is the purpose of this paper to investigate the  transition probability of the UDW detector in the presence of a non-relativistic particle as a starting point of such attempt, where the transition probability can be interpreted as the probability of finding the particle at the position of the detector. 

Working with a non-relativistic particle state allows us to compare the transition probability of the UDW detector with the well-understood probability density of the corresponding free Gaussian wave packet in the non-relativistic quantum mechanical description, which is proportional to the energy density of the field in the non-relativistic limit as we show in Sec.~\ref{secreview}. To keep the model simple with a focus on particle properties, we consider a massive scalar field living in the two-dimensional Minkowski spacetime, where we can consider the interaction to have a sharp switch-on and switch-off instead of introducing technical details of smearing the detector over time. We find that the total transition probability of the detector splits into the vacuum contribution and the matter contribution, and we show that the matter contribution gives a qualitatively similar description to the probability density of the particle. Such result indicates that our detector model can serve as a faithful field theoretic measurement model for the single particle detection. Unique features inherent to the detector model are found as the matter part contribution oscillates with the interaction time whose period is determined by the difference between the energy gap of the detector and the mass of the particle. Moreover, we observe that there is a strong resonance pattern for the transition probability when the energy gap of the detector is tuned to the mass of the particle.

The paper is organized as follows. In Sec.~\ref{secreview} we give a quick review of the field description for a non-relativistic particle and calculate its energy density.  In Sec.~\ref{secmain}, after introducing the UDW model and a quick review of the transition probability for the detector in the vacuum, we present our main results on the matter part contribution to the transition probability. Both analytical results for the detector coinciding with the particle and numerical results for more general scenarios are discussed. A comparison between the vacuum contribution and the matter part contribution has also been explored with different parameter choices. Sec.~\ref{secconlustion} gives concluding remarks of the paper. Throughout this paper, we use natural units $\hbar = c = 1$
and the metric of the two-dimensional Minkowski spacetime
has signature $(-,+)$.

\section{Single particle description in the two-dimensional Minkowski spacetime}
\label{secreview}
In this section, we briefly review the quantum field description of a non-relativistic particle.
Consider a free real scalar field $\phi(t, x)$ of mass $m$ in two-dimensional Minkowski spacetime, which satisfies the Klein-Gordan equation:
\begin{align}
(-\Box +m^2)  \phi(t, x) = 0,
\label{KGequation}
\end{align}
where $\Box \ce \partial_\nu \partial^\nu$ is the d'Alembertian operator. 
Solving this field equation and imposing the canonical quantization for the field, the expression of the field operator can be obtained as
\begin{align}
\phi\left(t,x\right) = \int \frac{d{k}}{\sqrt{(2\pi)^2 2\omega_{k}}} \left(a( k)e^{ik_{\mu}x^{\mu}}+ a^{\dag}( k) e^{-ik_{\mu}x^{\mu}} \right),
\end{align}
where $\omega_{k} = \sqrt{k^2 +m^2}$ is the energy of a single mode and the creation and annihilation operators satisfy the usual commutation rule:
\begin{align}
[a(k), a^\dag(k')] = \delta(k -k').
\end{align}

A non-relativistic particle localized at  $x_0$ and with momentum ${k}_0$ can be described by the initial field state \cite{milesinprogress}:
\begin{equation}
|\psi(0)\rangle =N \int \frac{dk}{\sqrt{(2\pi)^2 2\omega_{k}}} e^{-\frac{1}{2\sigma^2}({k}-{k}_0)^2-i{k} {x_0}}  a^{\dag}({k}) |0\rangle,
\label{inistateoffield}
\end{equation}
where $N$ is the normalization constant. The field state description of the particle is non-relativistic to a good approximation provided the characteristic radius satisfies $\sigma^{-1}\gg m^{-1}$ and the momentum ${k}_0$ satisfies $|{k}_0|\ll m$. Under these conditions, the normalization constant is approximately $|N|=2\pi^{1/4} \sqrt{m/\sigma}$.

A natural way to see that such an initial state provides a similar description to a free localized Gaussian wave packet with position $x_0$ and momentum $k_0$ in non-relativistic quantum mechanics is to compare the expectation value of the energy density  
$T_{00} = \frac{1}{2} \left[m^2\phi^2 +\dot{\phi}^2 + (\frac{\partial \phi}{\partial x})^2 \right]$ for the initial state with the time-dependent probability density of the corresponding wave-function. In the non-relativistic limit, the expectation value of the energy density approximately reduces to a simpler form:  $\langle T_{00} \rangle = m^2\langle \phi^2 \rangle$, where we have neglected the vacuum energy terms and also $(\frac{\partial \phi}{\partial x})^2$ term since it's proportional to $k_0^2$, which is small compared with $m^2$. Note that we also employed the fact that the expectation value of time derivative term reduces to $\frac{1}{2}m^2 \langle \phi^2 \rangle$ in such limit.

As derived in Appendix \ref{twopointsec}, the expectation value of $\phi^2$ for $|\psi(0)\rangle$ is given by
\begin{align}
&\langle \psi(0) |\phi\left(t, x\right)^2 |\psi(0)\rangle \nn \\
&= \frac{1}{m}\left[\frac{\sigma^2}{\pi\left(1+\left(\frac{\sigma^2 t}{m}\right)^2\right)}\right]^{\frac{1}{2}} \exp\left[-\sigma^2\frac{\left(x-x_0-\frac{k_0 t}{m}\right)^2}{1 + \left(\frac{\sigma^2 t}{m}\right)^2}\right],
\label{energydensity}
\end{align}
which coincides with the non-relativistic probability density up to a constant $m^{-1}$ (for quantum mechanical description, see Appendix B). From Eq.~(\ref{energydensity}), we see the variance of the energy density (probability density) grows with time $t$, indicating the particle state spreads spatially over time. 

\section{Transition probability of the Unruh-Dewitt detector}
\label{secmain}
The point like Unruh-Dewitt detector can be thought as a two-level system moving along some timelike spacetime trajectory $\mathsf{x_D(\tau)}$ where $\tau$ is the proper time of the detector. The Hilbert space of the detector is spanned by the ground state $|0_D\rangle$ and the excited $|1_D\rangle$ separated by an energy gap $\Omega$. The detector couples to the scalar field locally through the interaction Hamiltonian
\begin{align}
H_{int}(\tau) &= \lambda \chi \left(\tau \right)\Big(e^{ i\Omega \tau} \sigma^+  +  e^{- i\Omega \tau}\sigma^- \Big)  \phi\left[\mathsf{x}_D(\tau)\right], \label{InteractionHamiltonian}
\end{align}
where $\lambda$ is the coupling strength, $\sigma^+ =  |1_D\rangle \langle 0_D |$ and $\sigma^- = |0_D\rangle \langle 1_D |$ are ladder operators acting on the detector's Hilbert space, $\chi(t) $ is a compact switching function which controls the switch-on and switch-off moments of the interaction. 

In the following, we shall consider an inertial detector at rest at the origin of the coordinate system with its worldline given by the Minkowski coordinate\footnote{Without loss of generality, one can always go to the reference frame of the detector, provided that the relative speed between the particle and the detector is non-relativistic.}:
\begin{align}
\mathsf{x_D}(\tau) = (\tau, 0).
\label{trajec}
\end{align}
Such worldline has the simple interpretation that the particle position $x_0$ is also the separation distance between the particle and the detector.

\subsection{Transition probability in the vacuum background}
Supposing that the detector and the field states are initially prepared in the ground state $|0_D \rangle$ and the vacuum state $|0 \rangle$ before the interaction, the transition probability for the detector to jump to the excited state $|1_D\rangle$ after the interaction has ceased is given to the leading order of the coupling constant by \cite{louko2006often}
\begin{align}
P_{v} = \lambda^2 \int d\tau  d \tau' \chi(\tau) \chi(\tau') e^{-i \Omega \left(\tau-\tau'\right)} W_v\left(\mathsf{x}_D(\tau), \mathsf{x}_D(\tau')\right),
\label{vacuumP}
\end{align}
where $W_v\left(\mathsf{x}_D(\tau), \mathsf{x}_D(\tau')\right) \ce \langle 0 |\phi(\mathsf{x}_D(\tau)) \phi(\mathsf{x}_D(\tau'))| 0 \rangle$ is the pull back of the vacuum Wightman function to the detector's worldline. 

We remark that the vacuum Wightman function in general should be regarded a distribution on the spacetime and one usually needs to consider a smooth switching function $\chi(\tau)$ to cure the possible divergence in Eq.~(\ref{vacuumP}) in order to obtain unambiguous results for the transition probability. However, as a special case in the two-dimensional Minkowski spacetime for the free massive scalar field, the coincidence singularity of the vacuum Wightman function is only logarithmic \cite{decanini2008hadamard} and we can consider a  sharp switching function:
\begin{align}
\chi(\tau) = \Theta(\tau - \tau_i) \Theta(\tau_f - \tau),
\label{switchfunc}
\end{align}
where $\Theta(\tau)$ is the Heaviside step function and $\tau_i$ ($\tau_f$) indicates the switch-on (off) moment while $P_v$ remains well defined. Note that we have implicitly assumed that $\tau_f \geq \tau_i$, i.e., we always first switch on the interaction and then switch it off with a finite interaction time duration $\Delta \tau := \tau_f - \tau_i$.  

The pull back of the vacuum Wightman function for a massive scalar field in the two-dimensional Minkowski spacetime to the detector's worldline is \cite{takagi1986vacuum,toussaint2021detecting}
\begin{align}
W_v(\tau, \tau') = \frac{1}{2\pi} K_0(m[\epsilon +i(\tau - \tau')]),
\label{vacwightman}
\end{align}
where $K_0$ is the modified Bessel function of the second kind with limit $\epsilon \rightarrow 0^+$ understood. 

The transition probability $P_v$ then can be found as \cite{toussaint2021detecting} 
\begin{align}
&P_v =\nn \\
&-\frac{\lambda^2}{2m^2} \int_0^{\Delta \tilde{\tau}} du (\Delta \tilde{\tau} - u) [ J_0(u) \sin (\mu u) + Y_0(u) \cos(\mu u)],
\label{expressionofPv}
\end{align}
where $\Delta \tilde{\tau} = m (\tau_f - \tau_i)$, $\mu = \Omega/m$, $J_0$ and $Y_0$ are the Bessel's function of the first kind and the second kind.

\subsection{Transition probability in the presence of a particle}
Now we are ready to discuss the transition probability of the detector in the presence of a non-relativistic particle. Supposing that the field state is prepared as in Eq.~(\ref{inistateoffield}) at $\tau = 0$ with the detector in its ground state $|0_D\rangle$ and adopting the switching function $\chi(\tau)$ in Eq.~(\ref{switchfunc}) with $\tau_i \geq 0$, the transition probability for the detector to the leading order of the coupling strength is \cite{louko2006often}
\begin{align}
P_{p} =& \lambda^2   \int_{\tau_i}^{\tau_f} \int_{\tau_i}^{\tau_f} d\tau   d \tau'  e^{-i \Omega \left(\tau-\tau'\right)} \nn \\
&\times \langle \psi(0) | \phi(\mathsf{x}_D(\tau)) \phi(\mathsf{x}_D(\tau'))| \psi(0)\rangle.
\label{particleP}
\end{align}
The two-point function in Eq.~(\ref{particleP}) can be expressed as a sum of the vacuum contribution and the matter contribution (see derivation in appendix \ref{twopointsec}):
\begin{align}
\langle \psi(0) | \phi(\mathsf{x}_D(\tau)) \phi(\mathsf{x}_D(\tau'))| \psi(0)\rangle = W_v(\tau, \tau') + W_m(\tau, \tau'),
\end{align}
where
\begin{align}
&W_m(\tau, \tau') =  \frac{1}{2\sqrt{\pi}m\sigma} \frac{e^{-im (\tau - \tau') - \frac{k_0^2}{\sigma^2}}}{\sqrt{\left(\frac{1}{\sigma^2} +\frac{i\tau}{m}\right)  \left(\frac{1}{\sigma^2} -\frac{i\tau'}{m}\right)}} \nn\\
&\times \exp\left( \frac{\left(\frac{k_0}{\sigma^2} - ix_0\right)^2}{2\left( \frac{1}{\sigma^2} +\frac{i\tau }{m}\right)} +\frac{\left(\frac{k_0}{\sigma^2} +ix_0\right)^2}{2\left( \frac{1}{\sigma^2} -\frac{i\tau' }{m}\right)} \right) + \{  \tau \Longleftrightarrow \tau'\}.
\label{wightmanmatter}
\end{align}

Note that to reach Eq.~(\ref{wightmanmatter}), we have taken the non-relativistic limit approximation. Substituting Eq.~(\ref{wightmanmatter}) into Eq.~(\ref{particleP}), we then obtain the transition probability as a sum of the vacuum contribution and the matter part contribution:
\begin{align}
P_p &=  P_v + P_m \nn \\
&= P_v + \lambda^2   \int_{\tau_i}^{\tau_f} \int_{\tau_i}^{\tau_f} d\tau   d \tau'  e^{-i \Omega \left(\tau-\tau'\right)}W_m(\tau, \tau').
\label{particleprob}
\end{align}

The expression of $P_m$ is a complicated integral which does not admit an analytical form in general. In the following two subsections, we shall first discuss a special case of $x_0 = 0$ and $k_0=0$, where analytical results can be obtained and then we employ numerical methods to study the dependence of $P_m$ on other parameters.

\subsubsection{Analytical results}
In case of $x_0 = 0$ and $k_0 =0$, the point-like detector essentially overlaps with the particle and the matter part contribution to the two-point function simplifies to
\begin{align}
W_m(\tau, \tau') &=  \frac{1}{2\sqrt{\pi}m\sigma} \frac{e^{-im (\tau - \tau') }}{\sqrt{\left(\frac{1}{\sigma^2} +\frac{i\tau}{m}\right)  \left(\frac{1}{\sigma^2} -\frac{i\tau'}{m}\right)}} \nn\\
& + \{  \tau \Longleftrightarrow \tau'\}.
\label{Wmsp}
\end{align}
Substituting Eq.~(\ref{Wmsp}) into Eq.~(\ref{particleP}), we find
\begin{align}
P_{m} &= \frac{\lambda^2}{2\sqrt{\pi}m\sigma}   \int_{\tau_i}^{\tau_f} \int_{\tau_i}^{\tau_f} d\tau   d \tau'  e^{-i \Omega \left(\tau-\tau'\right)} \nn \\
&~~~\times \left(\frac{e^{-im (\tau - \tau') }}{\sqrt{\left(\frac{1}{\sigma^2} +\frac{i\tau}{m}\right)  \left(\frac{1}{\sigma^2} -\frac{i\tau'}{m}\right)}} 
+ \{  \tau \Longleftrightarrow \tau'\}\right) \nn \\
&= \frac{\lambda^2 m }{2 \sqrt{\pi} \sigma^3} \left( \Big \lvert I(\tau_f, \Omega) - I(\tau_i, \Omega) \Big\rvert^2  +\{\Omega \Longleftrightarrow -\Omega \} \right),
\label{Pmspecial}
\end{align}
where we have introduced function $I(\tau, \Omega)$ defined as
\begin{align}
I(\tau, \Omega) &\ce e^{\frac{2m(m+\Omega)}{\sigma^2}}\sqrt{1 + \frac{i \tau \sigma^2}{m}} \nn \\
&~~~\times E_{\frac{1}{2}} \left(\frac{m(m+\Omega)}{\sigma^2} + i(m + \Omega)\tau \right),
\label{defineIfunc}
\end{align}
with $E_{1/2}$ the exponential integral.

We now discuss properties of $P_m$ in Eq.~(\ref{Pmspecial}). From Eq.~(\ref{Pmspecial}) we see that $P_m$ is invariant under the change of $\Omega \rightarrow - \Omega$, indicating that the matter part contribution to the transition probability is the same for both the excitation and the de-excitation of the detector\footnote{If $\Omega$ is a negative quantity, $|1_D\rangle$ effectively becomes the ground state with $|0_D\rangle$ being the excited state.}. As a matter of fact, this conclusion also applies to particles with non-zero momentum $k_0$ and position $x_0$ since  $W_m(\tau, \tau')$ is a symmetrical function of $\tau$ and $\tau'$ and the transformation of $\Omega \rightarrow -\Omega$ in Eq.~(\ref{particleP}) amounts to the exchange of variables $\tau$ and $\tau'$ in the integral, which then gives the same result. We note that the vacuum Wightman function $W_v(\tau, \tau')$ is, however,  non-symmetrical.

A closer study of the function $I(\tau, \Omega)$ reveals more details on how $P_m$ depends on time $\tau$ and energy gap $\Omega$. In the long time limit $\tau_f \rightarrow +\infty$ (which corresponds to infinite interaction time duration $\Delta \tau \rightarrow +\infty$), $I(\tau_f, \Omega)$ approaches  zero asymptotically, resulting $P_m$ an initial time $\tau_i$ dependent quantity. This asymptotic property of the $I(\tau, \Omega)$ function also means that $P_m$ decreases as $\tau_i$ gets larger with a fixed interaction time duration, which is in agreement with the fact that the particle state spreads spatially over time with a decreasing energy/probability density. We note that the exponential integral has an oscillatory dependence on its imaginary component, and therefore $P_m$ also oscillates with interaction time duration $\Delta \tau$. For a positive value of $\Omega$, the first term in Eq.~(\ref{Pmspecial}) is dominated by the second term and the period of $P_m$ is approximately given by $T = 2\pi/( m- \Omega)$. Moreover, as we shall see in the following, there is a strong resonance effect at $\Omega = m$ where $P_m$ obtains its peak value.

\begin{figure}[t]
\begin{center}
\includegraphics[height = 2.6in]{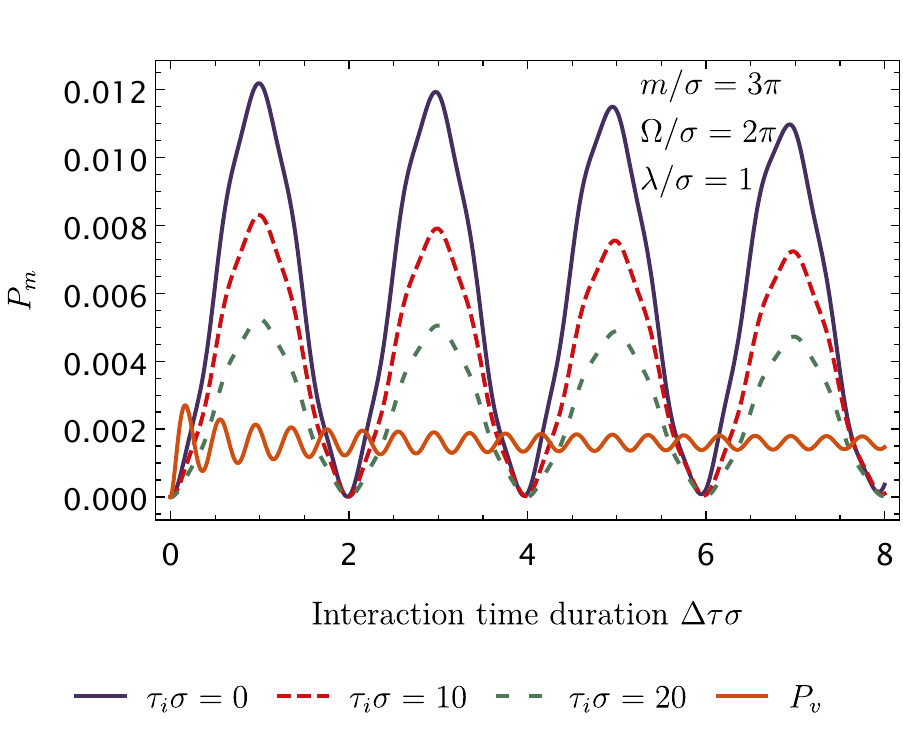} 
\end{center}
\caption{The matter part contribution to the transition probability $P_m$ and the vacuum part contribution $P_v$ are plotted as functions of the dimensionless interaction time duration $\Delta \tau \sigma$. We see that both $P_m$ and $P_v$ oscillate with the time duration and its peak value decreases gradually over time.}
\label{figpmspecialtime}
\end{figure}

\begin{figure}[t]
\begin{center}
\includegraphics[height = 2.6in]{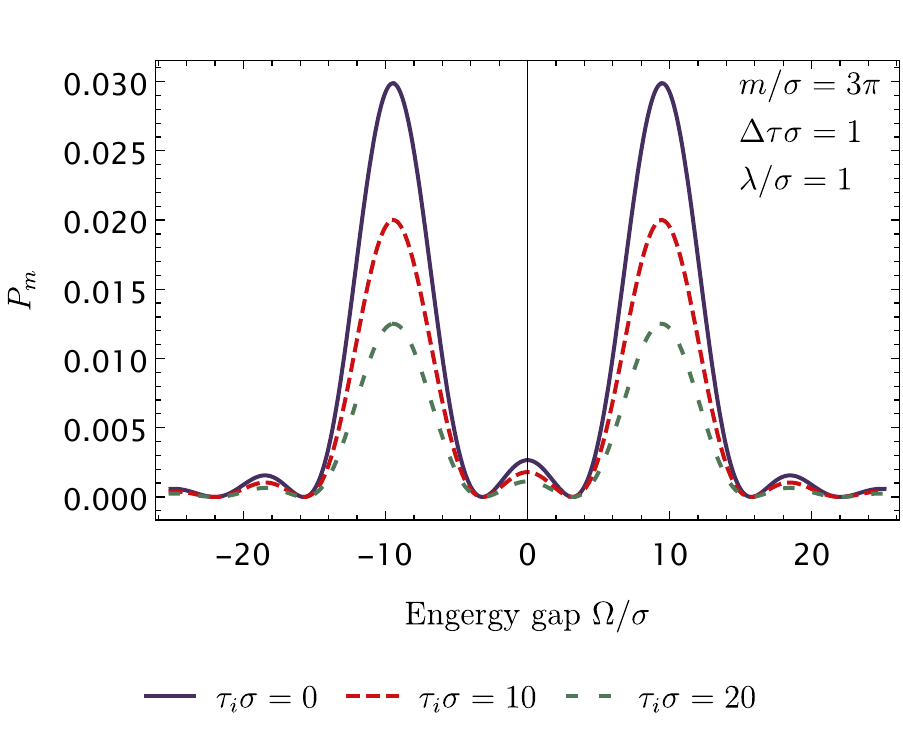} \\

\end{center}
\caption{The matter part contribution to the transition probability $P_m$ is plotted as a function of the dimensionless detector energy gap $\Omega/ \sigma$. We see a symmetrical dependence of $P_m$ on the energy gap with its peak values obtained in the resonance condition $\Omega =  \pm m$. }
\label{figpmspecialgap}
\end{figure}

\begin{figure}[t]
\begin{center}
\includegraphics[height = 2.6in]{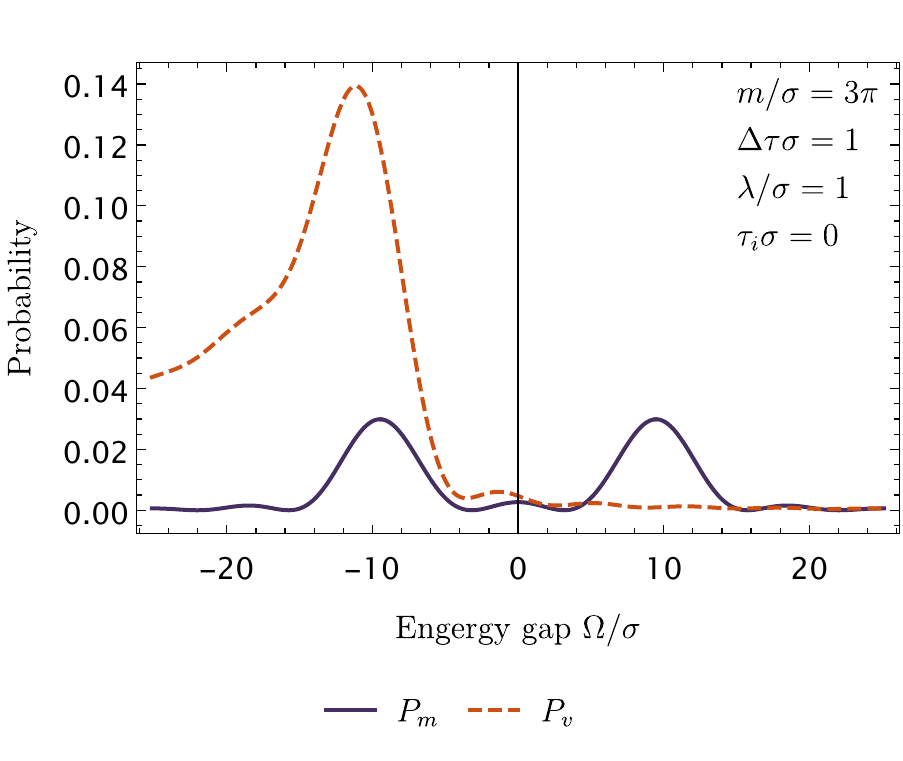} \\

\end{center}
\caption{The matter part contribution to the transition probability $P_m$ and the vacuum part contribution $P_v$ are plotted as functions of the dimensionless energy gap $\Omega/\sigma$. We see that $P_m$ dominates $P_v$ for positive value of the energy gap in the resonance region, and for negative value of the energy gap $P_v$ gets significantly larger than $P_m$ with its peak around $\Omega = -m$.}
\label{figpcomparisongap}
\end{figure}

We plot $P_m$ as a function of the dimensionless interaction time duration $\Delta \tau \sigma$ and the energy gap $\Omega/\sigma$ with different switch-on moments $\tau_i$ in Figs.~\ref{figpmspecialtime} and \ref{figpmspecialgap}. From Fig.~\ref{figpmspecialtime}, we see $P_m$ oscillates with a period approximately of $2\pi/(m/\sigma-\Omega/\sigma) = 2\sigma$ and its peak value gradually decreases over time as we remarked previously. Comparing different switch-on moments $\tau_i \sigma$, we see the transition probability gets smaller for larger values of $\tau_i \sigma$ with a fixed interaction time duration. For a comparison with the vacuum contribution, the dependence on the interaction duration of $P_v$ is also plotted in Fig.~\ref{figpmspecialtime}, and it can be seen that the amplitude of $P_v$ is much smaller than $P_m$ here. Fig.~\ref{figpmspecialgap} shows the symmetrical dependence of $P_m$ on the dimensionless energy gap $\Omega/\sigma$. Moreover, we see from Fig.~\ref{figpmspecialgap} that there is a strong resonance effect for $P_m$ when the energy gap of the detector is tuned to $\Omega =  \pm m$. Such resonance should come as no surprise since that the non-relativistic particle would have the same energy as the excited state of the detector in this case. Again, Fig.~\ref{figpmspecialgap} reveals a smaller $P_m$ with larger values of the starting time moment $\tau_i \sigma$ with other parameters fixed.

In Fig.~\ref{figpcomparisongap} we compare the dependence on the dimensionless energy gap $\Omega/\sigma$ for $P_m$ and $P_v$. It is seen that for positive value of $\Omega/\sigma$, $P_v$ is dominated by $P_m$ when the energy gap of the detector is close to the resonance condition $\Omega = m$, which is in agreement with Fig~\ref{figpmspecialtime}. However, for negative value of $\Omega/\sigma$, $P_v$ is significantly larger than $P_m$ with its peak around $\Omega = -m$ \cite{toussaint2021detecting}, indicating that the detector has much higher probability to de-excite  compared with the excitation rate in the presence of vacuum and it is less sensitive to the matter part contribution for the de-excitation.

We end this subsection with some more discussion on the resonance effect for $P_m$. Taking $m = \Omega$, the second term in Eq.~(\ref{Pmspecial}) is in fact ill-defined since $E_{1/2}(0)$ is formally infinite. This apparent infinity is due to the improper treatment of the integration in Eq.~(\ref{Pmspecial}). Taking $m = \Omega$ in the integral, we obtain
\begin{align}
P_m =&  \frac{\lambda^2 m }{2 \sqrt{\pi} \sigma^3} \Bigg[ \Big \lvert I(\tau_f, \Omega) - I(\tau_i, \Omega) \Big\rvert^2  \nn \\
&+ \bigg \lvert \sqrt{ 1- \frac{i \tau_f \sigma^2}{m}} - \sqrt{ 1- \frac{i \tau_i \sigma^2}{m}} \bigg \rvert^2 \Bigg]
\label{pmspecialresonance}.
\end{align}
The expression of $P_m$ in Eq.~(\ref{pmspecialresonance}) is, however, also problematical if one consider large difference between $\tau_f$ and $\tau_i$ as $P_m$ can gets larger than one. Such divergence implies that in the resonance condition, the first order result for the transition probability is invalid for long interaction time duration and one has to take into account the contribution from higher order terms. We note that such first order divergence is due to the stronger infrared divergence in the lower dimensional quantum field theory. The power of the denominator in Eq.~(\ref{Wmsp}) increases with the dimension of the spacetime, and therefore Eq.~(\ref{pmspecialresonance}) would converge in higher dimensional spacetimes.

\subsubsection{Numerical results}
\begin{figure}[h]
\begin{center}
\subfloat[]{\includegraphics[height = 3.1
in]{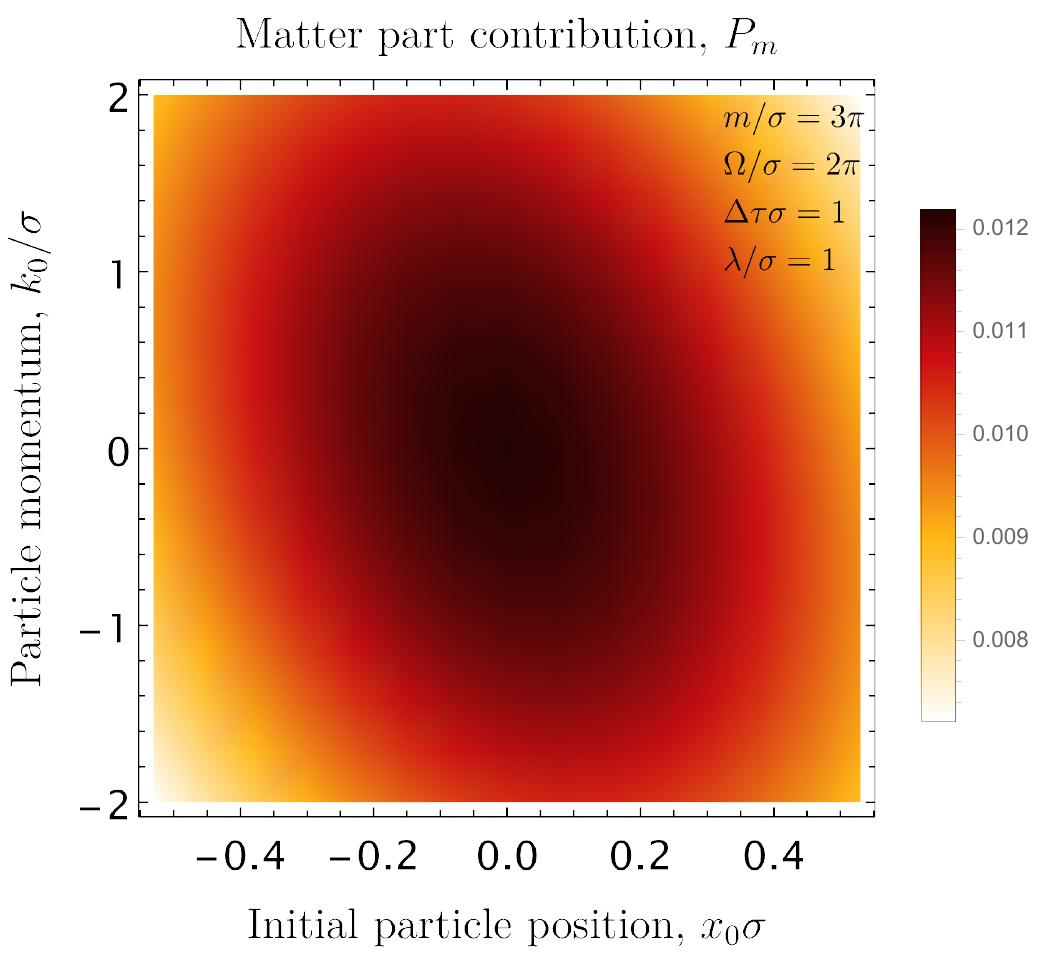}}
\\
\subfloat[]{\includegraphics[height = 2.6 in]{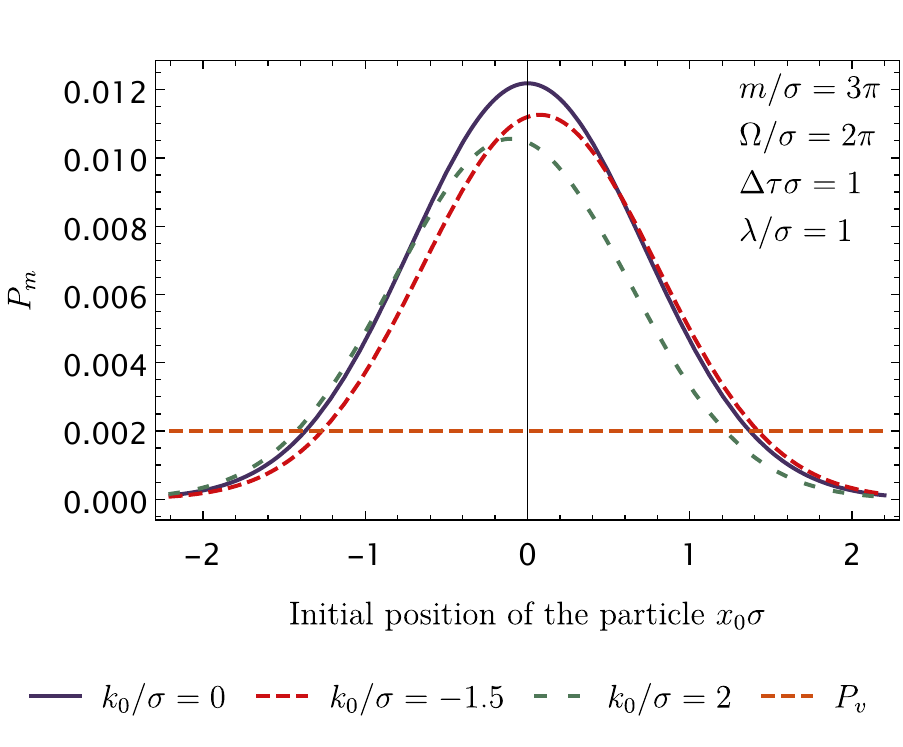} }
\end{center}
\caption{(a) The matter contribution to the transition probability $P_m$ is plotted as a function of the dimensionless initial particle position $x_0\sigma$ and the particle momentum $k_0/ \sigma$. We see the peak of the probability moves in the opposite directions in the position space as the momentum increases in the positive direction. (b) A more study of $P_m$ is plotted versus initial particle position $x_0 \sigma$ with different momentum $k_0/\sigma$. The switch-on moments in both (a) and (b) are $\tau_i \sigma = 0$.}
\label{figdensity}
\end{figure}

The integration for $P_m$ with non-zero initial position $x_0$ and momentum $k_0$ can hardly be evaluated analytically. With the dependence on time $\tau$ and energy gap $\Omega$ discussed in the previous subsection, we shall employ numerical methods in this subsection to focus on exploring the dependence of $P_m$ on $x_0$ and $k_0$. 

Some comments are in order here before we discuss the numerical plots. Similar to the energy/probability density in Eq.~(\ref{energydensity}), it can be seen from  Eq.~(\ref{wightmanmatter}) that $W_m(\tau, \tau')$ roughly decreases exponentially with the square of the particle position $x_0$, and therefore $P_m$ would also fall off exponentially with larger value of $x_0$. Furthermore, since $P_v$ is independent of the particle position $x_0$, in case of sufficiently large separation between the particle and the detector, the vacuum contribution would dominate the matter contribution.

Fig.~\ref{figdensity} displays the numerical plot of $P_m$.  Fig.~\ref{figdensity} (a) is a density plot of $P_m$ as a function of the dimensionless initial particle position $x_0 \sigma$ and the particle momentum $k_0/\sigma$, from which one can see that for each fixed value of the momentum $k_0/\sigma$, there is a corresponding peak of the probability in the position space. As the particle deviates from such peak position, $P_m$ falls off in both directions quickly. Furthermore, we see that as the momentum $k_0/\sigma$ increases in the positive direction, the peak of the $P_m$ in the position space moves in the opposite direction to the negative values for $x_0 \sigma$. Fig.~\ref{figdensity} (b) depicts a more detailed numerical study on $P_m $ as a function of the initial particle position $x_0 \sigma$ with different momentum $k_0/\sigma$ as well as the vacuum contribution to the transition probability $P_v$.  We see that as the detector sits sufficiently far from the particle, the vacuum contribution $P_v$ gets greater than the matter contribution $P_m$ and eventually dominates it. Moreover, it can be seen more clearly that for zero momentum,  $P_m$ falls off exponentially in both directions in a symmetrical fashion, and in case of the non-zero momentum for the particle, the peak of $P_m$ is shifted in the corresponding direction by a certain value as we have seen in Fig.~\ref{figdensity} (a). We remark that such behaviour agrees with the energy/probability density dependence in the phase space of the particle.  Intuitively one would expect that $P_m$ should be larger if the particle is moving towards the detector during the interaction time interval in contrast to the case when it's moving away from the detector since the average energy/probability density during the interaction time interval at the position of the detector is greater in the former case. 

\begin{figure}[h]
\begin{center}
\includegraphics[height = 2.6in]{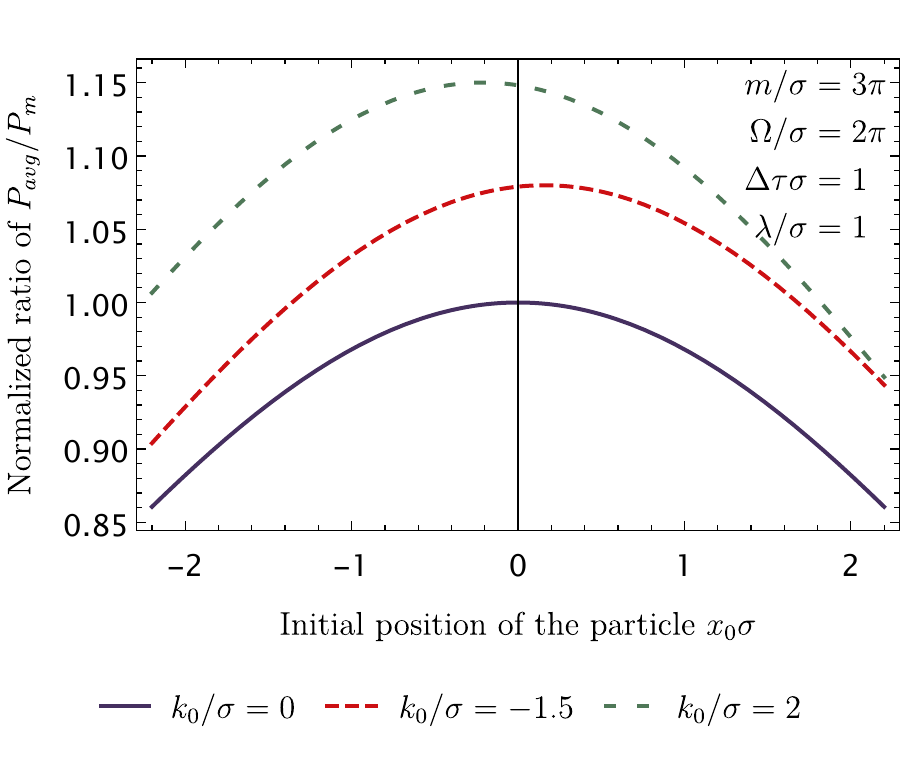} 
\end{center}
\caption{The normalized ration of $P_{avg}/P_{m}$ is plotted as a function of the dimensionless initial particle position $x_0 \sigma$ with different momentum $k_0/\sigma$. The normalization is taken such that $P_{avg}/P_m = 1$ for $x_0\sigma = 0$ and $k_0/\sigma = 0$.}
\label{figratio}
\end{figure}

However, the similarity between the non-relativistic probability density and $P_m$ should only be understood in a qualitative sense. To compare the matter part contribution to the transition probability of the detector with the non-relativistic probability density of the particle, we define the averaged probability density at the position of the detector as
\begin{align}
P_{avg} \ce \frac{m}{\tau_f - \tau_i} \int_{\tau_i}^{\tau_f} d \tau  \langle \psi(0) |\phi\left(\tau, 0\right)\phi\left(\tau, 0\right) |\psi(0)\rangle,
\end{align}
where we have used the fact that the expectation value of $\phi^2$ coincides with the non-relativistic probability density up to a constant $m^{-1}$. Fig.~\ref{figratio} shows the normalized ratio plot of $P_{avg}/P_{m}$ versus the initial particle position $x_0 \sigma$ with different momentum $k_0/\sigma$, where an implicit normalization constant has been taken such that $P_{avg}/P_m$ equals to $1$ for $x_0 = 0$ and $k_0 = 0$. It can be seen that these two quantities do not have a strict linear relationship and the matter part contribution $P_m$ decays slower over the separation distance between the detector and the particle compared with $P_{avg}$. 
\\
\\

\section{Conclusion and outlook}
\label{secconlustion}
In this work we studied in detail the transition probability of the UDW detector in the presence of a non-relativistic particle. We introduced an initial state of a massive scalar field that represents a non-relativistic particle and we calculated it's energy density which is shown to be proportional to the corresponding non-relativistic probability density. 

Coupling the UDW detector to such an initial field state, we found that the transition probability splits into the vacuum contribution and the matter part contribution. An analytical result for the matter part contribution is obtained in the special case when the particle coincides with the detector during the interaction. It was shown that the matter contribution oscillates with the interaction time duration, and with its peak gradually deceasing over time to its initial time dependent asymptotic value. The frequency of the oscillation is determined by the difference between the mass of the particle and the energy gap of the detector. When the mass equals to the energy gap, we found a strong resonance effect for the transition probability. The comparison between the vacuum contribution and the matter part contribution was performed and we found that for the excitation of the detector, the matter contribution would mostly dominate the vacuum contribution while for the de-excitation of the detector, the situation is reversed. We employed numerical methods to investigate the more general scenarios when the particle does not coincide with the detector and we found that the matter part contribution behaves similar to the averaged energy density of the particle at the position of the detector during the interaction. Such similarity, as we have checked, should only be understood in a qualitative sense. 

Although we have done the analysis in a two-dimensional flat spacetime, we expect that most properties of the matter contribution to the transition probability are still valid in higher dimensional spacetime as the two-point functions for a non-relativistic particle state share similar structures. Our work has paved the way for operationally investigating field properties in the presence of matter. It would be interesting to extend the analysis to either more general matter distribution scenarios (such as superposition or entangled excitation state) or different interaction types. In particular, it's worth investigating if there exists a type of the interaction between the detector and the field that reproduces exactly non-relativistic probability result. Finally, we notice that it's also interesting to explore how are the entanglement properties of the field influenced by the matter presence as seen by a pair of the UDW detectors, which we postpone to the future work.\\ \\

\begin{acknowledgments}
We would like to thank Miles Blencowe, Alexander Smith and Shadi Ali Ahmad for useful discussions and comments. This work is supported by the Dartmouth Teaching Fellowships. \end{acknowledgments}

\bibliography{main}

\onecolumngrid

\appendix
\section{Derivation of the two point function}
\label{twopointsec}

The purpose of this section is twofold. We shall first derive the expectation value of $\phi^2$ as in Eq.~(\ref{energydensity}) and then calculate the two point function as in Eq.~(\ref{wightmanmatter}).  

Using the expression of the initial field state in Eq.~(\ref{inistateoffield}) and sandwiching two field operators in between, we have:
\begin{align}
&\langle \psi(0) |\phi\left({x},t\right)\phi\left(x',t'\right) |\psi(0)\rangle =
|N|^2\int \frac{d{k_1}}{(2\pi)^{1/2}(2\omega_{k_1})^{1/2}} e^{-\frac{1}{2\sigma^2}({k_1}-{k}_0)^2-i{k_1} {r_0}}\int \frac{d{k_2}}{(2\pi)^{1/2}(2\omega_{k_2})^{1/2}} e^{-\frac{1}{2\sigma^2}({k_2}-{k}_0)^2+i{k_2} {r_0}} \nn \\
&~~~\times
 \int \frac{d{k}}{(2\pi)^{1/2}(2\omega_{k})^{1/2}} \int \frac{d{k^\prime}}{(2\pi)^{1/2}(2\omega_{k^\prime})^{1/2}} \langle 0 |a(k_2) \left( a(k) a^{\dag}(k^\prime) e^{ix^{\mu}k_{\mu}-i x'^{\mu}k_{\mu}^\prime}+a^{\dag}(k)a(k^\prime) e^{ix'^{\mu}k'_{\mu}-ik_{\mu} x^{\mu})} \right)a^{\dag}(k_1)|0\rangle.
\label{twopointfuncexpression}
\end{align}
where we have dropped odd multiples of creation/anihilation operators since they give vanishing result. Using the  commutation relation $[a(k),a^{\dag}(k')]=\delta(k-k')$, the expectation values of the operator products in Eq. (\ref{twopointfuncexpression}) can be simplified to 
\begin{align}
 \langle 0|a(k_2)a(k)a^{\dag}(k^\prime)a^{\dag}(k_1) |0\rangle = \delta(k-k^\prime) \delta(k_1 - k_2)+ \delta(k-k_1)\delta(k^\prime-k_2),
 \label{opeq1}
\end{align}
and
\begin{align}
\langle0|a(k_2)a^{\dag}(k)a(\mathbf{k}^\prime)a^{\dag}(k_1) |0\rangle = \delta(k_1-k^\prime) \delta (k_2-k).
\label{opeq2}
\end{align}
Substituting Eq.~(A2) and Eq.~(A3) into Eq.~( \ref{twopointfuncexpression}) we then have:
\begin{align}
 \langle \psi(0) |\phi\left({x},t\right)\phi\left(x',t'\right) |\psi(0)\rangle &=
|N|^2\int \frac{d{k_1}}{4 \pi \omega_{k_1}} e^{-\frac{1}{\sigma^2}({k_1}-{k}_0)^2}\int \frac{d{k}}{4 \pi \omega_{k}} e^{ik_{\mu} ( x - x')^{\mu}}\nn \\
&+ |N|^2\int  \frac{d{k_1}}{4 \pi \omega_{k_1}} e^{-\frac{1}{2\sigma^2}({k_1}-{k}_0)^2 - ik_1 x_0 +ik_{1\mu}x^{\mu}}    \int  \frac{d{k_2}}{4 \pi \omega_{k_2}} e^{-\frac{1}{2\sigma^2}({k_2}-{k}_0)^2 + ik_2 x_0 -ik_{2\mu}x'^{\mu}}  \nn \\
&+|N|^2\int  \frac{d{k_1}}{4 \pi \omega_{k_1}} e^{-\frac{1}{2\sigma^2}({k_1}-{k}_0)^2 - ik_1 x_0 +ik_{1\mu}x'^{\mu}}    \int  \frac{d{k_2}}{4 \pi \omega_{k_2}} e^{-\frac{1}{2\sigma^2}({k_2}-{k}_0)^2 + ik_2 x_0 -ik_{2\mu}x^{\mu}}.   
\end{align}
As can be easily checked, the first line of Eq.~(A4) is just the vacuum Wightman function of the scalar field.

We now first derive the energy density term in Eq.~(\ref{energydensity}). Setting $x = x'$ and $t = t'$, Eq.~(A4) reduces to
\begin{align}
 \langle \psi(0) |\phi\left({x},t\right)\phi\left(x,t\right) |\psi(0)\rangle &=
|N|^2\int \frac{d{k_1}}{4 \pi \omega_{k_1}} e^{-\frac{1}{\sigma^2}({k_1}-{k}_0)^2}\int \frac{d{k}}{4 \pi \omega_{k}} +2 |N|^2 \bigg \lvert \int  \frac{d{k_1}}{4 \pi \omega_{k_1}} e^{-\frac{1}{2\sigma^2}({k_1}-{k}_0)^2 - ik_1 x_0 +ik_{1\mu}x^{\mu}} \bigg \rvert^2   
\end{align}
The first integral corresponds to the infinite vacuum energy term which we shall ignore. To evaluate the second integral, we employ the non-relativistic approximation by expanding the $\omega_{k} t$ phase terms to second order in ${k}$ and making the approximation $\omega_{k} =m$ for the terms appearing in the denominators, the resulting approximate Gaussian integral is
\begin{align}
\langle \psi(0) |\phi\left(t, x\right)\phi\left(t, x\right) |\psi(0)\rangle = \frac{1}{m}\left[\frac{\sigma^2}{\pi\left(1+\left(\frac{\sigma^2 t}{m}\right)^2\right)}\right]^{\frac{1}{2}} \exp\left[-\sigma^2\frac{\left(x-x_0-\frac{k_0 t}{m}\right)^2}{1 + \left(\frac{\sigma^2 t}{m}\right)^2}\right],
\end{align}
which is Eq.~(\ref{energydensity}).

Next we calculate the pull back of the two-point function to the detector worldline which is given in Eq.~(\ref{trajec}). Replacing the operator $\phi\left({x},t\right)\phi\left(x',t'\right)$ by  $\phi\left(0,\tau \right)\phi\left(0,\tau'\right)$ in Eq.~(A(4)) and adopting the similar approximation methods, we have
\begin{align}
\langle \psi(0) | \phi\left(0,\tau \right)\phi\left(0,\tau'\right)| \psi(0)\rangle =W_v(\tau, \tau') + W_m(\tau, \tau'),
\end{align}
where $W_v(\tau, \tau')$ is given in Eq.~(\ref{vacwightman}) and $W_m(\tau, \tau')$ can be found as
\begin{align}
W_m(\tau, \tau') =  \frac{1}{2\sqrt{\pi}m\sigma} \frac{e^{-im (\tau - \tau') - \frac{k_0^2}{\sigma^2}}}{\sqrt{\left(\frac{1}{\sigma^2} +\frac{i\tau}{m}\right)  \left(\frac{1}{\sigma^2} -\frac{i\tau'}{m}\right)}}  \exp\left( \frac{\left(\frac{k_0}{\sigma^2} - ix_0\right)^2}{2\left( \frac{1}{\sigma^2} +\frac{i\tau }{m}\right)} +\frac{\left(\frac{k_0}{\sigma^2} +ix_0\right)^2}{2\left( \frac{1}{\sigma^2} -\frac{i\tau' }{m}\right)} \right) + \{  \tau \Longleftrightarrow \tau'\}.
\end{align}

\section{\label{sec:comparison}A free quantum particle description}
Consider a Gaussian wave packet state that describes a particle with position $x_0$ and momentum $k_0$:
\begin{align}
\Psi(x,t=0)= N \int d{k} e^{-\frac{1}{2\sigma^2}(k-k_0)^2+ik(x - x_0)} = N (2\pi\sigma^2)^{3/2} e^{-\frac{\sigma^2}{2}(x-x_0)^2} e^{ik_0 (x-x_0)}
\end{align}
where $N=(2\sigma \pi^{3/2})^{-3/2}$ is the normalization constant. The time evolution of the particle state can be obtained by solving the Schr\"{o}dinger equation for the free Hamiltonian $H = p^2/{(2m)}$, and we have
\begin{align}
\Psi({\mathbf{r}},t)=\left[\frac{\sigma}{\sqrt{\pi}\left(1+i t\sigma^2/m\right)}\right]^{1/2} \exp\left[-\frac{\sigma^2}{2}\frac{\left(x-x_0 -k_0 t/m\right)^2}{1+i\sigma^2 t/m} + i k_0 \left(x-x_0\right)-ik_0^2 t/{(2 m)}\right],
\label{freeparticlesoln}
\end{align}
from which one then finds probability density as
\begin{align}
|\Psi({\mathbf{r}},t)|^2 =\left[\frac{\sigma^2}{\pi\left(1+\left(\frac{\sigma^2 t}{m}\right)^2\right)}\right]^{1/2} \exp\left[-\sigma^2\frac{\left(x-x_0-{k_0 t}/{m}\right)^2}{1 + \left(\frac{\sigma^2 t}{m}\right)^2}\right].
\label{probdensityeq}
\end{align}
We see this result coincides with the expectation value of $\phi^2$ in Eq.~(\ref{energydensity}) up to a constant of $m$.

\end{document}